\documentclass[a4paper]{article}
\usepackage{Odyssey2022}
\usepackage{epsfig,amssymb,amsmath}
\usepackage{amsmath,graphicx}
\usepackage{subfig}
\usepackage{multirow}
\usepackage{enumitem}
\usepackage{url}

\ninept

\title{Multimodal Emotion Recognition using Transfer Learning from Speaker Recognition and BERT-based models}

\name{Sarala Padi$^{1}$, Seyed Omid Sadjadi$^{1}$, Dinesh Manocha$^{2}$, and Ram D. Sriram$^{1}$}

\address{$^{1}$NIST, ITL, Gaithersburg, MD \\ $^{2}$University of Maryland, College Park, MD}

\begin{document}
%
\maketitle
\begin{abstract}

Automatic emotion recognition plays a key role in computer-human interaction as it has the potential to enrich the next-generation artificial intelligence with emotional intelligence. It finds applications in customer and/or representative behavior analysis in call centers, gaming, personal assistants, and social robots, to mention a few. Therefore, there has been an increasing demand to develop robust automatic methods to analyze and recognize the various emotions. In this paper, we propose a neural network-based emotion recognition framework that uses a late fusion of transfer-learned and fine-tuned models from speech and text modalities.    More specifically, we i) adapt a residual network (ResNet) based model trained on a large-scale speaker recognition task using transfer learning along with a spectrogram augmentation approach to recognize emotions from speech, and ii) use a fine-tuned bidirectional encoder representations from transformers (BERT) based model to represent and recognize emotions from the text. The proposed system then combines the Resnet and BERT-based model scores using a late fusion strategy to further improve the emotion recognition performance. The proposed multimodal solution addresses the data scarcity limitation in emotion recognition using transfer learning, data augmentation, and fine-tuning, thereby improving the generalization performance of the emotion recognition models. We evaluate the effectiveness of our proposed multimodal approach on the interactive emotional dyadic motion capture (IEMOCAP) dataset. Experimental results indicate that both audio and text-based models improve the emotion recognition performance and that the proposed multimodal solution achieves state-of-the-art results on the IEMOCAP benchmark.

\end{abstract}
%
\section{Introduction}\label{sec:intro}
As machines and computer-based applications and interactions are continuously progressing, and becoming a part of our daily
lives, natural human-computer interaction (HCI) has become increasingly critical~\cite{review-schuller, ser-intro-richard}. As a result, understanding and responding to the emotional state of individuals is a necessary step in natural HCI~\cite{ser-intro-beale}. Apart from this, there are other applications where automatic detection and classification of human emotion plays a critical role and can be useful in psychological and nano-physiological studies of human emotional expression.      Similarly, emotion information plays an important role in automatic tutoring systems, call center services, gaming, personal assistants, to mention a few. Another potential application  can be alerting drivers by detecting their stress or anger levels that could lead to major accidents and impair their driving capabilities~\cite{ser_intro_cowie, ser-intro-beale}.

There are a number of modalities with which humans convey their emotions. Examples include speech, text, facial expressions, hand gestures, and so on. Although there are scenarios where facial expressions could be a preferred and more effective
way of communicating emotions, speech and text data are abundantly available and more conveniently captured as compared to
other modalities. In particular, speech plays a crucial role in conveying the emotional state of a human in the form of prosody
and/or paralinguistic context. For speech emotion recognition (SER), traditionally, machine learning (ML) models were developed using engineered features such as mel-frequency cepstral coefficients (MFCC), Chroma-based features, pitch, energy, entropy, and zero-crossing rate~\cite{intro-mfcc-ververidis, ser-dnn_han, intro-ml-kwon}, to mention a few. However, the performance of such ML models depends on the type and diversity of the features used. Although it remains unclear which features correlate most with various emotions, the research is still ongoing to explore additional features and new algorithms to model the dynamics of feature streams representing human emotions. On the other hand, deep learning-based models can directly learn the task-relevant features from spectrograms or raw waveforms~\cite{intro-wav-spect-yang, ser-spect-satt, ser-spect-ma, acm_cnngao}, thereby obviating the need for extracting a large set of engineered features [12]. Recent studies have proposed the use of convolutional neural network (CNN) models combined with long short-term memory (LSTM) built on spectrograms and raw waveforms, showing improved SER performance \cite{ser-spect-satt, ser-spect-ma, ser-spect-yenigalla, Attentive_michael, CNN-RNN-schuller, spectral-mirsamadi, sarma2018emotion}. However, building such complex systems requires large amounts of labeled training data. Also, insufficient labeled training data can potentially make the models overfit to specific data conditions and domains, resulting in poor generalization performance on unseen data conditions.

Another important modality for conveying and capturing emotions is text. To represent and model the textual data for solving natural language processing (NLP) related tasks, different approaches have been developed. The word and sentence embeddings have emerged as the most effective representation and have shown breakthroughs in improving the performance of deep learning models for various NLP applications. The variations of such embeddings include Word to Vector (Word2Vec)~\cite{word2vec-mikolov2013efficient}, Global Vectors (GloVe)~\cite{ glove-pennington2014}, and bidirectional encoder representations from transformers (BERT)~\cite{ bert-devlin2018bert}. Although these methods provide a compact representation of textual data, there are some limitations and challenges in effectively and efficiently leveraging these methods for various tasks. The main challenge is that building such models requires large amounts of data for the task at hand. To overcome this problem, transfer learning and fine-tuning the available pretrained models have shown promise. Nonetheless, the pretrained models are typically trained and tuned for out-of-domain tasks. Accordingly, the tweaked features may not perform well on target tasks such as emotion recognition, where the vocabulary used to extract the emotion is different from that used for the source task.

This paper presents a multimodal framework for emotion recognition from speech and text using transfer learning and fine-tuning of speaker ResNet and BERT-based models. This paper leverages the advantages of a BERT-based model combined with acoustic feature representations to improve  emotion recognition performance. In particular, we use a BERT model to not only  extract sentence embeddings from the text but also to fine-tune the model to improve the emotion recognition performance. In addition, to address the data scarcity limitations in speech-based emotion recognition, we use a transfer learning approach combined with a spectrogram augmentation strategy. Specifically, we re-purpose a ResNet~\cite{resnet} model developed for speaker recognition using large amounts of speaker labeled data and use it as a feature descriptor for SER. The model includes a statistics pooling layer that enables the processing of variable-length segments without a need for truncation. Also, we increase the training data size by generating more data samples using spectrogram augmentation\cite{spechaug}. The proposed framework effectively leverages the complementary information from text and speech modalities and combines the scores produced to improve emotion recognition performance. We evaluate the effectiveness of our proposed systems on the interactive emotional dyadic motion capture (IEMOCAP) dataset \cite{iemocap} using speech-only, text-only, and multimodal settings and show that the fusion of the two complementary modalities results in state-of-the-art emotion recognition performance.

\section{Related Work}

In recent years, deep neural network based systems have shown a tremendous success in recognizing the emotions from the speech signal [24, 16, 25]. Specifically, techniques such as bidirectional LSTMs (BLSTM)~~\cite{huang2016emotion, spectral-mirsamadi, table-ramet2018context, table-tripathi} and time-delay neural networks (TDNN)~\cite{wu2021bert}, which can effectively model relatively long contexts compared to their DNN counterparts, have been successfully applied for SER. Nevertheless, as discussed previously, the lack of large amounts of carefully labeled data for building complex models for emotion classification remains a main challenge in SER~\cite{acm_albanie2018emotion}. To address this, data augmentation methods are used to generate additional training data by perturbing, corrupting, mimicking, and masking the original data samples to enable the development of complex ML models.     For example, \cite{sarma2018emotion, ser-gan-Aggelina, pappagari2020xvector} applied signal-based transformations such as speed perturbation, time-stretch, pitch shift, as well as added noise to original speech waveforms. One disadvantage of these approaches is that they require signal-level modifications, thereby increasing the computational complexity and storage requirements of the subsequent front-end processing. They can also lead to model overfitting due to potentially similar samples in the training set, while random balance can potentially remove useful information~\cite{ser-gan-Aggelina}. Furthermore, retaining performance on relatively clean conditions has shown to be a challenge while using signal-based transformations~\cite{pappagari2020xvector}.

Another effective way to address data scarcity is transfer learning~\cite{ser-tl-review-feng, gideon2017progressive, boateng2020speech}. Transfer learning can leverage the information and knowledge learned from one related task and domain to another, as long as the input remains the same. Several studies have have shown that transfer learning approaches outperform prior methods in recognizing emotions even for unseen scenarios, individuals, and conditions~\cite{ser-tl-gideon}. It was also shown that transfer learning increases the feature learning abilities which leads to improved SER performance~\cite{ser-tl-ghosh,ser-tf-latif,ser-tf-schuller,ser-tf-song}. However, transfer learning methods have not been fully explored and analyzed for emotion recognition. Particularly, it is unclear whether and how ML models trained for other data-rich speech applications such as speaker recognition would perform for SER. We recently proposed an SER framework based on transfer learning from speaker recognition along with data augmentation~\cite{padi2021improved} which achieved promising improvements in SER performance on the IEMOCAP. In this study, we explore the impact of full network fine-tuning (as opposed to only training the fully connected classification layers) combined with transfer learning on SER performance.

As for text based approaches, word embeddings have emerged as the de facto representation in many natural language processing tasks. The most widely used word embedding models are Word2Vec~\cite{word2vec-mikolov2013efficient} and GloVe~\cite{glove-pennington2014}. Both these models are unsupervised and have shown great success for various NLP tasks including sentiment analysis, document indexing, and topic model analysis. However, a major limitation of these models is that the word order is not taken into account which leads to a loss of syntactic and semantic understanding of words in sentences. In addition, fine-tuning these word embeddings for different target tasks and applications is not feasible. To circumvent these limitations, BERT models are used for extracting representations from text data to capture the context. Several studies have successfully combined BERT-based embeddings with the speech-based representations to improve the emotion recognition performance~\cite{table-pepino2020fusion, tabel-makiuchi2021multimodal, wu2021bert}. However, BERT-based embeddings are context dependent and the representations extracted using the pretrained models may not capture the salient domain-dependent information for emotion recognition. This study investigates fine-tuning of a BERT model for text-based emotion recognition

In recent years, ML approaches have been explored where models are trained using multiple modalities and scores are combined by adding an extra layer (i.e., a meta learning layer) either by concatenating (stacking) the embeddings or scores, or using a weighted sum of the embeddings or scores~\cite{intro-majumder2019dialoguernn, endtoend-Tzirakismultimodal, DL-Ranganathanmultimodal}. Multimodal processing and recognition have shown promise in many practical machine learning applications~\cite{intro-sebe2005multimodal,intro-haq2011multimodal}. Since spoken data is composed of audio and text content, prior studies have also investigated the combination of acoustic features/representations and linguistic information for emotion recognition. Particularly, multimodal emotion recognition models have been developed from audio, video, images, or textual data.     The models have been combined either at the feature/representation level (aka the early fusion) or at the decision/score level (aka the late fusion)~\cite{table_multi_Haiyang,wu2021bert,table-pepino2020fusion}. In this study, we investigate the latter within a multimodal emotion recognition framework using speech and text models.

\section{Proposed Multimodal Framework}\label{sec:system}
\begin{figure*}
\centering
 \includegraphics[scale=.55, clip, trim=30mm 25mm 30mm 30mm]{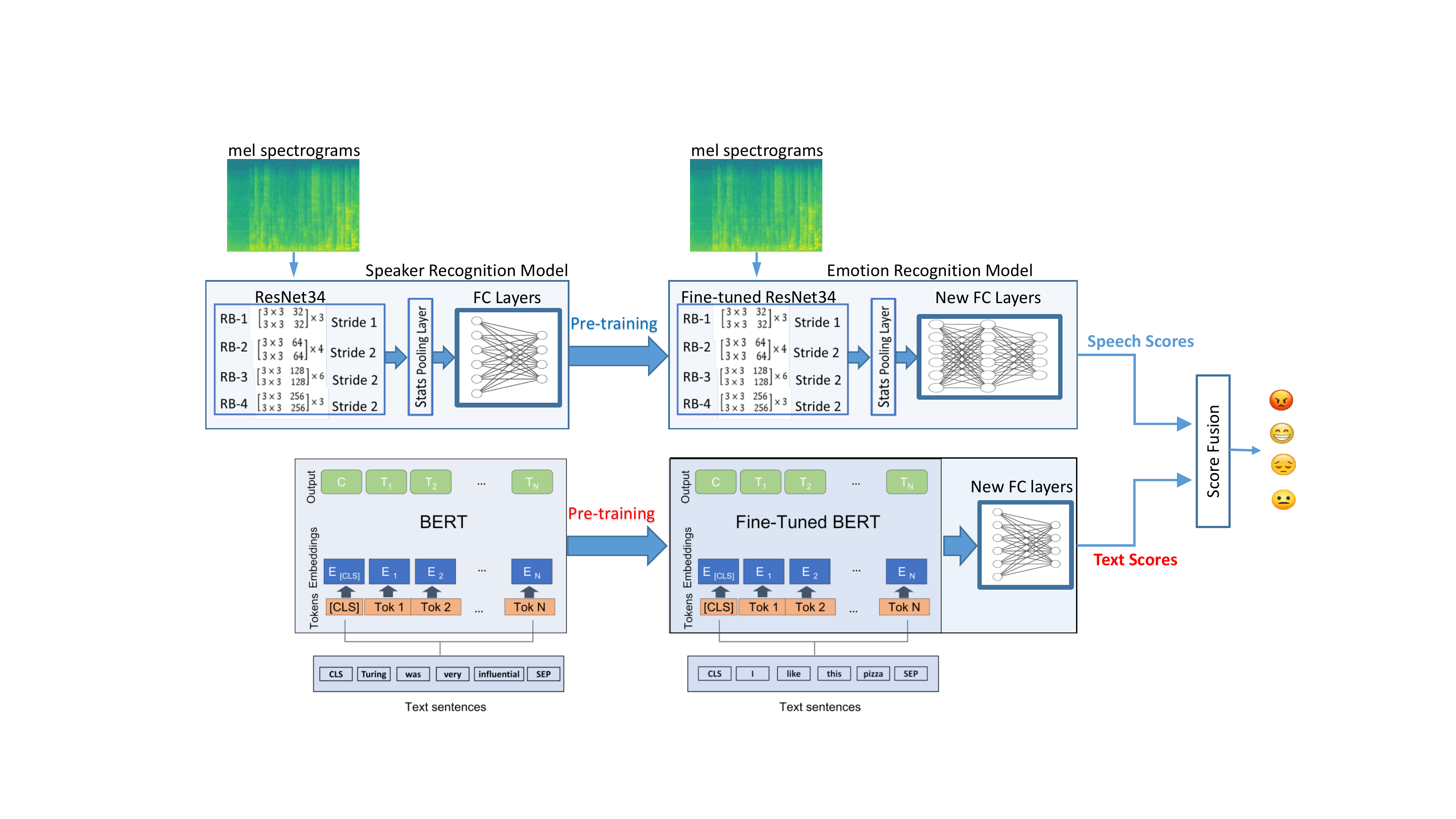}
   \caption{\it Block diagram of the proposed multimodal emotion recognition framework.}
\label{fig:method}
\end{figure*}

Figure~\ref{fig:method} shows the block diagram of the proposed multimodal emotion recognition framework. The following subsections provide a brief description of the speech and text-based models and systems used in this framework.   

\subsection{Text system}

\subsubsection{BERT}
BERT is a language model trained on large amounts of text data that has achieved state-of-art results on many NLP tasks. Because of its bidirectionality and use of transformers with the attention mechanism and positional encoding~\cite{ transformers-vaswani2017attention}, the BERT model learns the contextual information from input sentences. The BERT models are context-dependent. So, to use these models for downstream tasks, fine-tuning is required to generate the vectors/embeddings based on the word context. Unlike GloVe, it represents the input as subwords and learns sub-word embeddings (as opposed to whole word embeddings). By using the sub-word representation, the BERT approach overcomes the out of vocabulary limitation of character (ELMO)~\cite{elmo-peters2018deep} and word-based representations (GloVe or Word2Vec)~\cite{word2vec-mikolov2013efficient, glove-pennington2014}.

\subsubsection{Transfer learning}


In the BERT setup, preprocessing of the input text is performed to generate tokens and are mapped to indices. As shown in Figure~\ref{fig:method}, there are two additional tokens called the [CLS] classification token and the [SEP] separate segment token, which are appended at the beginning and end of each sentence, respectively. First, the embedding layer receives a list of input tokens and generates word embeddings for each token by adding positional embeddings to maintain the word order in a sequence.  Next, the output of the embedding layer is fed to the multi-head self-attention sublayer to model temporal dependencies. Attention masks are used to exclude the paddings in sequences from attention weight calculations. Finally, we extend the BERT-base model by adding a new classification head, including randomly initialized fully connected (FC) layers with a softmax to get the prediction probabilities for the various emotions. We also fine-tune the network parameters on the training set for the emotion recognition task.

\subsection{Speech system}\label{subsec:system_speech}

Although SER systems traditionally used a large set of low-level time- and frequency-domain features to capture and represent the various emotions in speech, in recent years, many state-of-the-art SER systems have used complex neural network models that learn directly from spectrograms or even raw waveforms. In this study, we build and explore an end-to-end ResNet~\cite{resnet} based system using log-mel spectrograms as input features. ResNet models originally, developed for computer vision applications \cite{resnet}, have recently gained interest in speech applications such as speaker recognition \cite{zeinali2019but}. We extract high-resolution spectrograms to enable the model to learn the spectral envelope and the coarse harmonic structures for the various emotions.

\subsubsection{Transfer learning}

As noted previously, transfer learning is an ML method where a model initially developed for one task or domain is re-purposed, partly or entirely, for a different but related task/domain. It has recently gained interest in SER \cite{ser-tl-review-feng}. In this study, we re-purpose a model initially developed for speaker recognition to serve as a feature descriptor for SER. We first train a ResNet34 model on large amounts of speaker-labeled audio data, and then we replace the FC layers of the pre-trained model with new randomly initialized FC layers. Finally, we 1) only re-train the new FC layers for an SER task on the IEMOCAP dataset ( \textit{linear probing}~\cite{head2toe}), and 2) re-train the FC layers and fine-tune the convolutional layers using two different learning rates. As shown in Figure~\ref{fig:method}, the proposed system employs a statistics pooling layer \cite{snyder2018xvector} that aggregates the frame-level information over time and reduces the sequence of frames to a single vector by concatenating the mean and standard deviation computed over frames. The convolutional layers in the ResNet model work at the frame level, while the FC layers work at the segment level.



\subsubsection{Spectrogram augmentation}



There has been recent success in applying a computationally efficient data augmentation strategy, termed spectrogram augmentation, for speech recognition tasks \cite{spechaug}. The spectrogram augmentation technique generates additional training data samples by applying random time-frequency masks to spectrograms to mitigate the over-fitting issue and improve the generalization of speech recognition models. Motivated by promising results seen with the spectrogram augmentation in the speech recognition field, we augment the training data using spectro-temporally modified versions of the original spectrograms. Because the time-frequency masks are applied directly to spectrograms, we can conveniently apply the augmentation on the fly by eliminating the need to create and store new data files. Similar to the approach taken in \cite{spechaug}, we consider two policies, conservative and aggressive, to generate spectrogram augmentations for SER. The parameter settings for the two spectrogram augmentation policies are like \cite{padi2021improved}.

\subsection{Multimodal Fusion}
To leverage the complementary information captured by the two modalities, we use a weighted average of the scores ($S_{speech}$ and $S_{text}$) obtained from the speech and text models as follows:
\begin{equation} \label{eq:1}
    S_{fusion} = w_{1} \times S_{speech} + w_{2} \times S_{text}.
\end{equation}

The scores are calculated by applying the natural logarithm to outputs of the softmax layer (i.e., class posteriors). Here, $w_{1}$ and $w_{2}$ are fixed weights assigned to the text and speech modalities, respectively. We constrain the weights to sum to 1, i.e., $w_2 = 1 - w_1 $. The weights determine the degree to which each modality contributes to the final score. For emotion classification, we select the emotion category with the highest score. 

We investigate two approaches for finding the fusion parameters. In the first approach, we directly use the raw scores from each modality and search for the best setting for $w_1$ on a hold-out set ($w_2$ is simply $1 - w_1$). In the second approach, we first pre-processed the scores for each modality to have zero mean and unit variance as follows:

\begin{equation} \label{eq:2}
    S_n = \frac{S - \mu_S}{\sigma_S},
\end{equation}
where the normalization statistics $\mu_S$ and $\sigma_S$ are calculated on a hold out set. The final fusion score is then obtained by averaging the normalized scores using $w_1=w_2=0.5$ (i.e., equal weight fusion).

\section{Experimental Setup} \label{sec:exp}

\subsection{Dataset}
We evaluate the effectiveness of the proposed multimodal emotion recognition system on the IEMOCAP dataset \cite{iemocap}, which contains improvised and scripted multimodal dyadic conversations between actors of opposite genders. It has 12 hours of speech data from 10 subjects and is pre-segmented into shortcuts. It includes nine categorical emotions and 3-dimensional labels. Our experiments include categorical emotions and speech segments with majority labeling, where at least two annotators agree in the annotation labeling. To replicate the experimental protocols used in many prior studies, we consider four categorical emotions: ``angry’’, ``happy’’, ``neutral’’, and ``sad’’ where we merged ``happy’’ and ``excited’’ into one category and the number of examples per category is $1103$, $1636$, $1708$, and $1084$, respectively. Table~\ref{tab:statistics} summarizes the data statistics in the IEMOCAP dataset for the experimental setup considered in this study. 

The IEMOCAP dataset comprises five sessions, and the speakers in the sessions are non-overlapping. Therefore, there are ten speakers in the dataset, i.e., five female and five male speakers. To conduct the experiments in a speaker-independent fashion, we use a leave-one-session-out (LOSO) cross-validation strategy, which results in 5 different train-test splits/folds. For each fold, we use the data from 4 sessions for training and the remaining session for model evaluation. Since the dataset is multi-label and imbalanced, besides the overall accuracy and termed weighted accuracy (WA), we report the average recall over the different emotion categories, termed unweighted accuracy (UA), to present our findings. To understand and visualize the performance of the proposed system within and across the various emotion categories, we compute and report confusion matrices for the experiments. We use similar setups for the speech-only, text-only, and combined model.

\begin{table}[ht!]
\caption{\it Data statistics for the various emotion classes in the IEMOCAP dataset.}
    \centering
    \begin{tabular}{l|c}
    \hline\hline
         Emotion & Number of Examples   \\        \hline         \hline
          Angry & 1103 \\
         Happy+Excited & 1636 \\
         Neutral & 1708 \\
         Sad & 1084 \\\cline{1-2}
         \textbf{Total} & \textbf{5531} \\ \hline\hline
    \end{tabular}
    
    \label{tab:statistics}
\end{table}


\subsection{Configuration}\label{subsec:imp}
\subsubsection{Speech system}
For speech parameterization, we extract high resolution 128-dimensional log-mel spectrograms from $25$~ms frames at a $100$~Hz frame rate (i.e., every $10$~ms). For feature normalization, we apply a segment level mean and variance normalization\footnote{No voice activity detection (VAD) is applied before feature normalization because it was found to be detrimental to SER performance.} which is not ideal because we applied a typical normalization at the recording/conversation level. We have found that normalizing the segments using statistics computed at the conversation level significantly improves the SER performance on the IEMOCAP. However, this violates the independence assumption for the speech segments and did not consider in this study. For the front-end processing, including feature extraction and feature normalization, we use the NIST speaker and language recognition evaluation (SLRE) \cite{sadjadi2020sre, sadjadi2018lre} toolkit. While training the model, we select $T$-frame chunks using random offsets over original speech segments, where $T$ is randomly sampled from the set $\{150, 200, 250, 300\}$ for each batch. We apply signal padding for speech segments shorter than $T$ frames. While evaluating the model, we feed the entire duration of the test segments because the statistics pooling layer enables the model to consume variable-length inputs.

As noted previously, the proposed end-to-end SER system uses a pre-trained ResNet34 model built on a speaker recognition task. We train the ResNet34 model with a $512$-dimensional embedding layer on millions of speech samples from over 7000 speakers available in the VoxCeleb corpus~\cite{nagrani2020voxceleb}. To build the speaker recognition model, we apply the same front-end processing described above to extract high-resolution log-mel spectrograms from VoxCeleb data. We conduct experiments using models with and without transfer learning and spectrogram augmentation. For each original speech segment, we generate and augment two spectro-temporally modified versions according to the augmentation policies defined in \cite{padi2021improved} for both speaker and speech emotion recognition systems during training. We also evaluate our models with and without the statistics pooling layer to study their impact on the emotion recognition task. We use a categorical cross-entropy loss as the objective function to train the models. The number of channels in the first block of the ResNet model is set to $32$. Pytorch\footnote{https://github.com/pytorch/pytorch} is used for model implementation, the stochastic gradient descent (SGD) optimizer with momentum ($0.9$), and a batch size of $32$. For transfer learning with frozen convolutional weights, an initial learning rate of $10^{-1}$ is used. To fine-tune the convolutional layers, we use a learning rate of $10^{-3}$. The learning rate remains constant for the first $8$ epochs and is halved for every other epoch. We use parametric rectified linear unit (PReLU) activation functions in all layers (except for the output) and utilize layer-wise batch normalization to accelerate the training process and improve the generalization properties of the model.

\setlength{\tabcolsep}{.5mm}
\renewcommand{\arraystretch}{1.}
\begin{table}[t]
\caption{\it Performance comparison of our proposed approach with prior methods that use the LOSO strategy for experiments (Audio, Text, and Audio+Text) on the full IEMOCAP dataset (i.e., both the improvised and scripted portions).  Abbreviations: A-Angry, H- Happy, N-Neutral, E-Excited, H+E: Happy and Excited merged. Bold letters indicate the best-performing model. Blanks (--) indicate unreported values.}
\label{tab:fusion_comp}
\centering
\begin{tabular}{l|l|c|c} 
 \hline\hline
 {Modality}&{Approach} & {UA [\%]} & {WA [\%]} \\  \hline\hline
  \multirow{8}{*}{Text}   &Haiyang Xu et.al~\cite{table_multi_Haiyang}    &60.3       &54.8\\
                               &Recurrent encoder~\cite{yoon2018multimodal}     & 63.5      &--\\ 
                               &BERT+self-attention~\cite{wu2021bert}   &58.53      &59.20\\ 
                               &Glove+self-attention~\cite{wu2021bert}         &61.27       &62.27\\ 
                               &BERT~\cite{table-pepino2020fusion} &55.2  &--\\ 
                               & BERT+CNN~\cite{tabel-makiuchi2021multimodal} &66.1   &67.0\\\cline{2-4}
                               &{\bf Proposed}   &{\bf 70.33}  &{\bf 70.24}\\\hline
    
    \multirow{19}{*}{Speech}& BLSTM+transfer learnig \cite{ser-tl-ghosh} & 51.86 & 50.47 \\
                               & VGG19+GAN augmentation \cite{ser-gan-Aggelina} & 54.6 & -- \\
                               &Recurrent encoder~\cite{yoon2018multimodal}    &54.6   &--\\ 
                               & CNN+multi-task learning \cite{Attentive_michael} & -- & 56.10 \\
                               &BLSTM+attention \cite{spectral-mirsamadi} & 58.8 & 63.5 \\
                               & CNN+transfer learning \cite{neumann2019attentive}           & 59.54 & -- \\
                               & ResTDNN+attention \cite{wu2021bert} & 61.32 & 60.64 \\
                               
                               &LSTM+attention \cite{table_multi_Haiyang}   &63.4   &57.4\\
                               
                               &TDNN+attention~\cite{wu2021bert} & 60.64      &61.32\\ 
                               &wav2vec-base \cite{ table-sp-yang-superb} &63.43   &--\\
                               &wav2vec-large \cite{table-sp-yang-superb} &65.64      &--\\
                               &HuBERT-base \cite{table-sp-yang-superb} & 64.92      &--\\
                               &HuBERT-large \cite{table-sp-yang-superb} &67.62   &--\\
                               &WavLM-base~\cite{ table-sp-chen-wavlm}    &65.94         &--\\
                               &WavLM-large~\cite{ table-sp-chen-wavlm}    &70.03   &--\\ 
                               \cline{2-4}
                               &{\textbf{Proposed} (w/o fine-tuning)} &{\bf 64.28}	&{\bf 63.74}\\
                               &{\textbf{Proposed} (w/ fine-tuning)} &{\bf 65.97}	&{\bf 65.4}\\\hline
    \multirow{6}{*} {\parbox{1cm}{Speech \\      + \\       Text}}&Recurrent encoder~\cite{yoon2018multimodal}        &71.8       &--\\ 
                                 &LSTM+attention+Concat \cite{table_multi_Haiyang}          & 70.4      &69.5\\
                                 &Audio+GloVe~\cite{wu2021bert}          &65.53       &66.43\\ 
                                 &BERT+CNN Fusion~\cite{table-pepino2020fusion}  & 65.1   &--\\
                                 &Wav2vec+CNN+BERT~\cite{tabel-makiuchi2021multimodal} &73.0    &73.5\\\cline{2-4} 
                                 &{\textbf{Proposed} ($w_1=0.94$)}  &{\bf 76.07}	&{\bf 75.76}\\
                                 &{\bf Proposed ($w_1=w_2=0.5$)}  &{\bf 75.97}	&{\bf 75.01}\\\hline\hline
\end{tabular}
\end{table}

\subsubsection{Parameter settings for text model}
The most commonly used pre-trained BERT models are BERT-base and BERT-large. The former has 12 encoders with 12 bidirectional self-attention heads, while the latter has 24 encoders with 16 bidirectional self-attention heads. These models were trained on unlabeled data collected from the BooksCorpus (800 million words) and English Wikipedia (2500 million words). In our experiments, we use a pretrained BERT-base model, specifically the BERT-base-uncased model with 12-layered transformer blocks with each block containing 12-head self-attention layers and $768$ hidden outputs. This model has 110 million parameters in total. We use randomly initialized fully connected layers with $768\times4$ parameters along with softmax, and fine-tune the model for emotion recognition on the IEMOCAP using a categorical cross-entropy loss as the objective function, an Adam optimizer with a learning rate of $2\times10^{-5}$, and a batch size of $32$. We use the Pytorch framework for the BERT model training.

\begin{figure*}[t]
\begin{center}
 \subfloat[\small {Text (BERT with fine-tuning)}]{\includegraphics[scale=0.5]{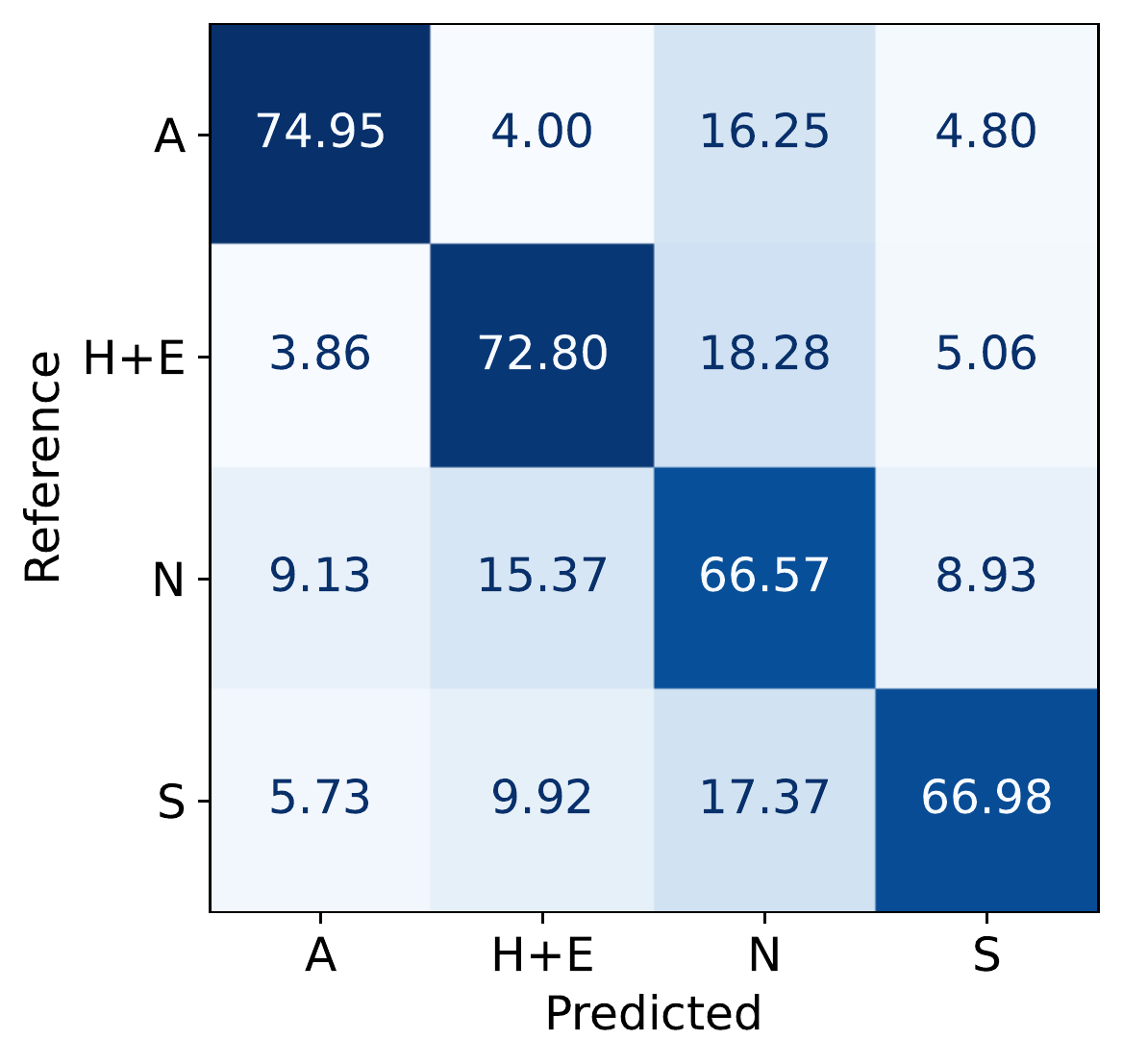}}
 \subfloat[\small {Speech (Resnet34 with fine-tuning)}]{\includegraphics[trim=.7cm 0cm 0cm 0cm, clip=true, scale=0.5]{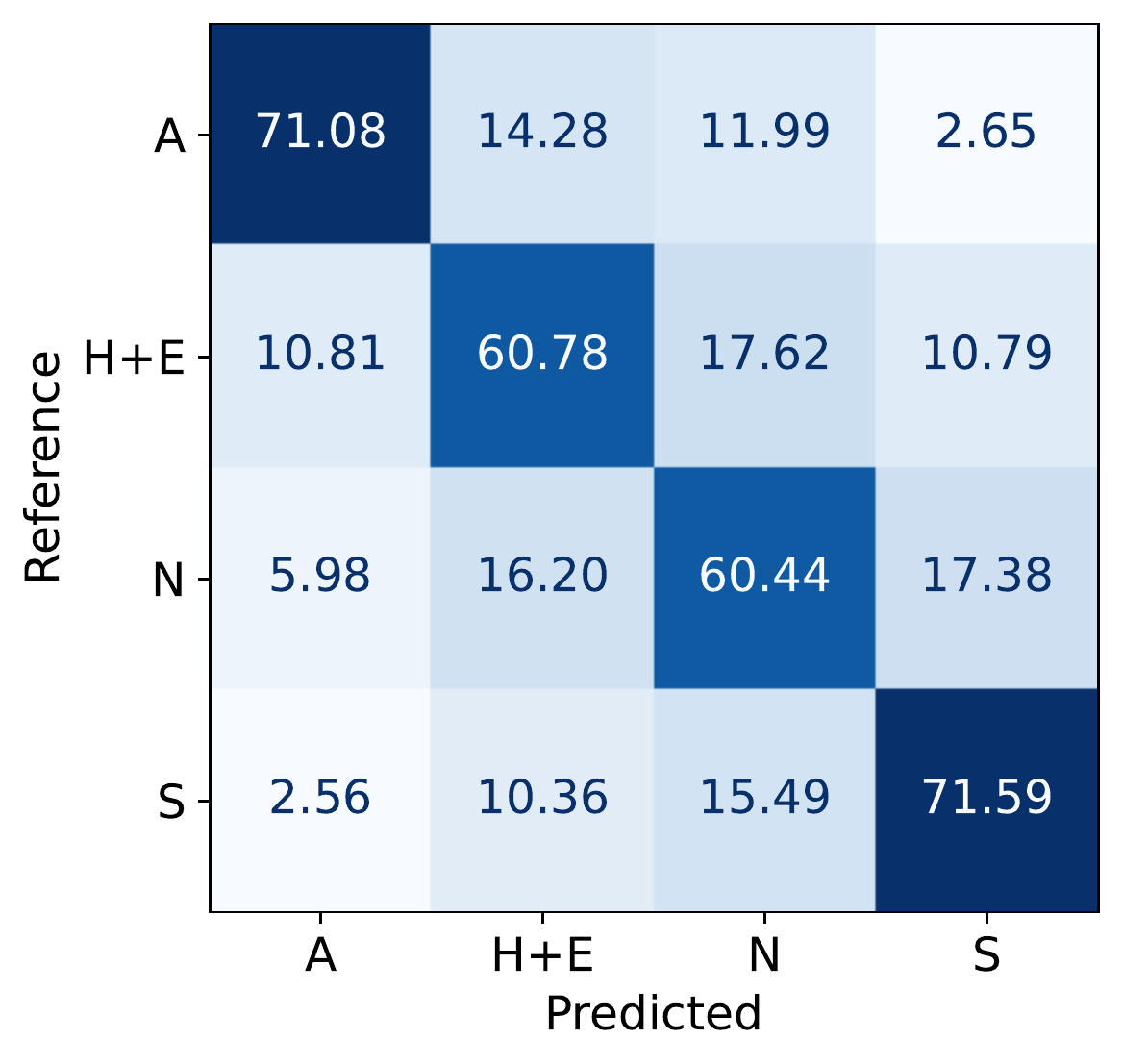}}
 \subfloat[\small {Multimodal fusion}]{\includegraphics[trim=.7cm 0cm 0cm 0cm, clip=true, scale=0.5]{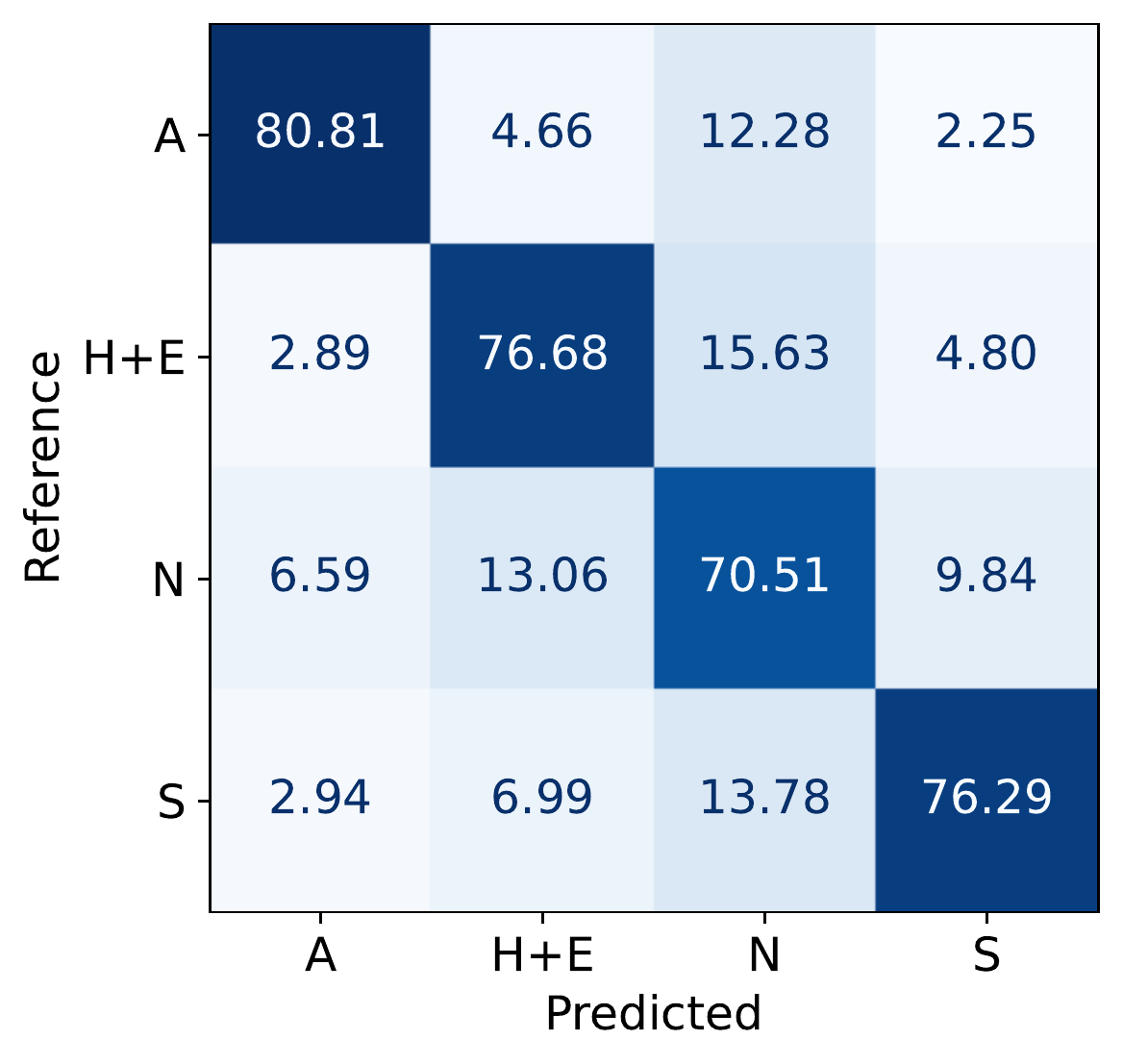}}
\end{center}
   \caption{\it Confusion matrices of the proposed approach for the speech-only, text-only, and multimodal experiments conducted in this study using the LOSO strategy. Abbreviations: A-angry, E-excited, H-happy, N-neutral, S-Sad, and H+E- Happy and Excited merged. Note that, we reported the average of performance metrics (WA \& UA) over the five session-wide folds as the result. }
\label{fig:confusion}
\end{figure*}

\section{Performance Evaluations}
\subsection{Results}

Table~\ref{tab:fusion_comp} presents the performance comparison of the proposed speech-only, text-only and multimodal systems with several
prior approaches for the experimental setup described in Section~\ref{sec:exp}. The fine-tuned BERT model outperforms prior text-based emotion recognition methods on the IEMOCAP with 70.33\% UA and 70.24\% WA, respectively. The fine-tuned ResNet34 model described in Section~\ref{subsec:system_speech} performs favorably compared to all SER approaches considered in this paper, except for HuBERT-large and WavLM-large with nearly 300 million parameters that use head-to-toe fusion of embeddings from 24 encoder layers. The proposed SER system comfortably outperforms a system that uses 384 engineered features~\cite{table-tarantino2019self}. We refer the reader to \cite{padi2021improved} for the results from the ablation study of the various components (i.e., transfer learning, data augmentation, statistics pooling) of the proposed SER system. In case of multimodal emotion recognition from speech and text, our proposed system outperforms prior methods and achieves state-of-the-art results using both fusion strategies. We note that combining the speech and text-based emotion recognition systems provides remarkable improvements with approximately 6\% absolute improvement compared to the best performing unimodal system (76.07\% vs 70.33\% UA).


\subsection{Error Analysis}
To visualize the performance of the proposed system within and across different emotion categories, confusion matrices for the
two modalities as well as their fusion are shown in Figure~\ref{fig:confusion}. It is observed from Figure~\ref{fig:confusion}(a) and (b) that the system confuses the ``happy'' class (H+E) with the ``neutral'' class (N) quite often, while performing the best on the ``angry'' emotion (A). This is consistent with observations reported in other studies on IEMOCAP~\cite{Attentive_michael, yoon2018multimodal}. Our informal listening experiments confirm that the ``happy'' and ``neutral'' classes are indeed confusable emotion pairs in the IEMOCAP dataset. Combining the Resnet34 and BERT model scores further improves the performance. Moreover, we observe that multimodal fusion reduces the confusion between ``Neutral'' (N) and ``Happy'' (H+E) emotions, performing best for ``angry'' class and achieving accuracies of over 70\% for the other three emotion categories.

\subsection{Cross comparisons}

\setlength{\tabcolsep}{1.5mm}
\renewcommand{\arraystretch}{1.}
\begin{table}[ht!]
\caption{\it Cross comparison of our proposed approach with prior methods  that use the LOSO strategy for experiments (speech- and text-only) on the full IEMOCAP dataset. `M' indicates million, Bold letters indicate the best-performing model and blanks (--) indicate unreported values. }
\label{tab:cross_comp}
\centering
\begin{tabular}{l|l|c|c} 
 \hline\hline
                {Method}                &{modality}   & {UA [\%]} & {Parameters} \\  \hline\hline
wav2vec-Base\cite{ table-sp-yang-superb}  &Audio       &63.43  &95.04M\\ 
wav2vec-Large\cite{ table-sp-yang-superb}         &Audio      &65.64  &317.38M \\
HuBERT-Base\cite{table-sp-yang-superb}            &Audio        &64.92 &94.68M \\
HuBERT-Large \cite{table-sp-yang-superb}          &Audio       &67.62  &316.61M \\
WavLM-Large\cite{ table-sp-chen-wavlm}            &Audio       &70.03  & 316.62M \\
\hline
\textbf{Proposed}    &Audio  &{\bf 65.97}   &\textbf{21.5M} \\\hline
BERT+attention\cite{wu2021bert}           &Text &58.53  &    --\\ 
BERT+CNN \cite{tabel-makiuchi2021multimodal} &Text &66.1   &--\\\hline
\textbf{Proposed}                        &Text &\textbf{70.33}  & \textbf{110M}\\\hline
\end{tabular}
\end{table}
All studies referenced in  Table \ref{tab:fusion_comp} adopt the LOSO strategy to conduct experiments on speech, text, and their fusion. There are other related studies in the literature that either only use the improvised portion of the IEMOCAP dataset or performed experiments using single fold as opposed to 5-fold cross validation \cite{ser-aug-caroline, ser-spect-satt,sarma2018emotion}. On the other hand, in our experiments, we use both the improvised and scripted portions of the IEMOCAP, which is approximately twice the size of the improvised portion alone. Because the experimental setups and the amount of data used for model training and evaluation in those studies are different than ours, we have not included them in Table ~\ref{tab:fusion_comp} for comparison. The emotion recognition performance on the improvised portion is known to be better than that on the full dataset (e.g., see \cite{table-tripathi, table-ramet2018context, table-tarantino2019self, ser-tl-ghosh, Attentive_michael}).


The latest HuBERT-base results reported in \cite{table-sp-yang-superb} uses a weighted average of embeddings from all transformer layers to achieve 67.62\% UA on the IEMOCAP task. However, the first version\footnote{https://arxiv.org/pdf/2105.01051v1.pdf} of the paper reported 62.94\% UA only using the embeddings from the last layer. This is more comparable to our proposed SER system with 65.97\% UA. Furthermore, from Table 3, we can notice that the HuBERT-large and WavLM-large models each have more than 300 million parameters as compared to our ResNet34 model with nearly 20 million parameters. Although HuBERT-large and WavLM-large models outperform our SER system, these models are computationally expensive and require large amounts of data for self-supervised training.


Similarly for the text modality, our fine-tuned BERT model outperformed prior methods. Note that although \cite{wu2021bert} reported a performance of 71.22\% UA, the experimental setup uses a context window [-3,3] including 3 preceding and succeeding sentences to generate the emotion prediction score for each sentence. This violates the independence assumption for the segments, hence it is not considered for comparison in this study. In addition, although this setup may be applicable to the IEMOCAP where context is inferred from the data, it may not generalize well to other datasets.




\section{Conclusions}
In this study, we presented a multimodal emotion recognition framework using transfer learning and fine-tuning of pre-trained
speaker recognition and BERT models. Specifically, we repurposed a pre-trained ResNet model from speaker recognition that was trained using large amounts of speaker-labeled data. Further, we fine-tuned the pretrained ResNet34 model to extract features from high-resolution log-mel spectrograms to improve the speech emotion recognition performance. In addition, we adopted a spectrogram augmentation technique to generate additional training data samples by applying random time-frequency masks to log-mel spectrograms to mitigate overfitting and improve the generalization of emotion recognition models. We also explored a BERT-based emotion recognition system and fine-tuned the pretrained weights to improve the emotion recognition performance. We further improved
the emotion recognition performance by fusing the complementary information available from speech and text modalities. We evaluated the proposed system using speech-only, text-only and multimodal settings and compared the performance of our system against that of several prior state-of-the-art studies. The proposed system consistently provided competitive performance across the two modalities as well as their fusion, achieving state-of-the-art results using text and multimodal settings. Results from this study show that models trained for data-rich applications such as speaker recognition and language modeling can be re-purposed using transfer learning to improve the emotion recognition performance under data scarcity constraints. In the future, we plan to extend our work to other datasets and other languages. We also plan to explore more efficient alternatives to network fine-tuning such as head-to-toe probing [51].

\section{Disclaimer}
	
These results presented in this paper are not to be construed or represented as endorsements of any participant's system, methods, or commercial product, or as official findings on the part of NIST or the U.S. Government.



\bibliographystyle{IEEEtran}
\bibliography{sample}

\end{document}